\documentclass[twocolumn,superscriptaddress,showpacs,floatfix,preprintnumbers, nofootinbib,hyperref]{revtex4} 
\usepackage{graphicx}
\usepackage{rotating}
\usepackage{epsfig}
\usepackage{bm}
\usepackage[usenames,dvipsnames]{xcolor}

\usepackage{color}

\def\he4{$^4$He}
\def\h2{$^2$H}

\newcommand{\lesssim}{\,\rlap{\lower3.7pt\hbox{$\mathchar\sim$}}
\raise1pt\hbox{$<$}\,}

\begin{document}

\title{{\color{ForestGreen}  Multi-azimuthal-angle effects in self-induced supernova  neutrino flavor conversions \\ without axial symmetry 
}}

\author{Alessandro Mirizzi} 
\affiliation{II Institut f\"ur Theoretische Physik, Universit\"at Hamburg, Luruper Chaussee 149, 22761 Hamburg, Germany}


\begin{abstract}
The flavor evolution of neutrinos emitted by a supernova (SN) core is strongly affected by 
the refractive effects associated with  the  neutrino-neutrino interactions in the deepest
stellar regions. Till now, all  numerical studies  have assumed the 
axial symmetry for the  ``multi-angle effects'' associated with the neutrino-neutrino interactions.
Recently, it has been pointed out in~\cite{Raffelt:2013rqa} that removing this assumption, a new multi-azimuthal-angle (MAA) instability
 would emerge in the flavor evolution of the dense SN neutrino  gas in addition to the one caused by multi-zenith-angle (MZA) effects. 
Inspired by this result, for the first time we numerically solve the non-linear neutrino propagation equations in SN, 
introducing
the azimuthal angle as angular variable in addition to the usual zenith angle. 
We consider simple energy spectra with an excess of  $\nu_e$ over $\bar\nu_e$.  
We find that even  starting with a complete axial symmetric neutrino emission, the MAA effects
would lead to significant flavor conversions in normal mass hierarchy, in cases otherwise stable under the only 
MZA effects.
The final outcome of the flavor conversions, triggered by the MAA instability, 
depends on the initial asymmetry between $\nu_e$ and $\bar\nu_e$ spectra. 
If it is sufficiently large, final spectra would show an ordered behavior with spectral swaps and splits. 
Conversely, for small flavor asymmetries flavor decoherence among angular modes develops, affecting the flavor evolution
also in inverted mass hierarchy. 
\end{abstract}


\maketitle

\section{Introduction}

Flavor conversions of  supernova neutrinos are
a fascinating problem involving refractive effects
that not only depend on the ordinary matter  
background~\cite{Matt,Dighe:1999bi}, but also on the neutrino fluxes themselves. Indeed,  
neutrino-neutrino interactions provide a non-linear term in the  
equations of motion~\cite{Pantaleone:1992eq, Sigl:1992fn,McKellar:1992ja} that causes  
collective flavor conversions~\cite{Qian:1994wh,Samuel:1993uw,  
  Kostelecky:1993dm, Kostelecky:1995dt, Samuel:1996ri, Pastor:2001iu,  
  Pastor:2002we, Sawyer:2005jk}.
Since few years~\cite{Duan:2005cp,Duan:2006an,Hannestad:2006nj} it has  realized that in the SN context these  
self-induced effects give rise to qualitatively new  
phenomena
(see, e.g.,~\cite{Duan:2010bg} for a recent review). The main consequence of 
this unusual type of  flavor transitions is 
an exchange of the spectrum of the electron species $\nu_e$ (${\bar\nu}_e$) with
the non-electron ones  $\nu_x$ (${\bar\nu}_x$) in certain energy intervals, giving
 rise to interesting spectral features~\cite{Duan:2006an,Hannestad:2006nj,Duan:2010bg,Fogli:2007bk,Fogli:2008pt, Raffelt:2007cb, 
Raffelt:2007xt, Duan:2007bt, Duan:2008za, Gava:2008rp, Gava:2009pj,Dasgupta:2009mg,%
Friedland:2010sc,Dasgupta:2010cd,Mirizzi:2010uz}. 
These flavor
exchanges are called ``swaps'' marked by the ``splits'',
which are the boundary features at the edges of each
swap interval.

In this context, one of the main complication in the simulation of the flavor evolution
is that the flux of neutrinos emitted from a 
 supernova core is far from isotropic. The current-current nature of the weak-interaction Hamiltonian implies that the interaction energy between neutrinos of 
momenta ${\bf p}$ and ${\bf q}$ is proportional to 
$(1-{\bf v}_{\bf p} \cdot {\bf v}_{\bf q})$, where ${\bf v}_{\bf p}$
is the neutrino velocity~\cite{Qian:1994wh,Pantaleone:1992xh}. In a non-isotropic medium this velocity-dependent term would not average to zero, producing a different 
refractive index for neutrinos propagating on different trajectories. 
This is the origin of the so-called ``multi-angle effects''~\cite{Duan:2006an}, which hinder the maintenance    of the coherent oscillation 
behavior for different neutrino modes~\cite{Duan:2006an,Raffelt:2007yz,Fogli:2007bk,EstebanPretel:2007ec,Sawyer:2008zs}.
All the numerical simulations of supernova neutrino flavor conversions have characterized
the multi-angle effects in the  so called 
 ``bulb model''~\cite{Duan:2006an} in which neutrinos are assumed spherically emitted 
at the neutrino-sphere at a radius $r=R$.~\footnote{The effect of non-trivial $\nu$ zenith angle distributions at the neutrinosphere 
has been discussed in~\cite{Mirizzi:2011tu,Mirizzi:2012wp}.}
Global spherical symmetry was also assumed  in the neutrino propagation, 
reducing to an axial symmetry of the multi-angle effects along the
radial line-of-sight (polar axis). 
Therefore, at  any radius $r > R$ along the polar axis, neutrinos with
 different momenta ${\bf p}$ (characterized by $|{\bf p}| = E$) were identified
 by the only zenith angle at the emission $\vartheta_R$. 
 
 The underlying assumption in the bulb model is that small deviations 
with respect to a perfect axial symmetry,   would not significantly affect
the  neutrino flavor evolution.  
This unjustified assumption has been questioned in a recent paper~\cite{Raffelt:2013rqa}, in which the authors investigated the effect
of relaxing the axial symmetry,   performing a linearized stability analysis of the SN neutrino equations of motion,
 introducing the azimuthal angle of neutrino
propagation $\varphi$ as explicit variable.
 The stability
analysis, as described in~\cite{Banerjee:2011fj}, would allow one to determine the possible onset of the flavor conversions, seeking for an exponentially
growing solution of the eigenvalue problem, associated
with the linearized equations of motion for the neutrino
ensemble.
Surprisingly it has been found that even assuming
a completely spherically symmetric neutrino emission at the neutrinosphere, without forcing the 
neutrino ensemble to follow the axial symmetric solution, a new ``multi-azimuthal-angle'' (MAA)
 instability would emerge in addition to the one caused by
the ``multi-zenith-angle'' (MZA) effects.  
In particular, considering simple neutrino ensembles with only $\nu_e$ and $\bar\nu_e$, the instability
has been found in normal mass hierarchy (NH,  ${\Delta m^2_{\rm atm}} = m_3^2-m_{1,2}^2>0$), where the system
would have been stable imposing   the  perfect axial symmetry. 
Subsequently, the role of these instability   has been clarified in~\cite{Raffelt:2013isa}, where it has been shown  with  simple toy models
that,  introducing small deviations in the assumed symmetries of the initial conditions
in a dense $\nu$ gas, new instabilities would be triggered.

 Once  this new instability has been found in the dense SN neutrino gas, 
 numerical simulations of the flavor evolution are mandatory 
in order to characterize how the instable system evolves. 
At this regard, the   main aim of this work is to perform a  numerical study of the SN neutrino flavor evolution
including both  neutrino zenith and azimuthal propagation angles. 
It it known that the development of self-induced flavor conversions in SNe, as well as their dependence on the
neutrino mass hierarchy, is crucially dependent on the
flux ordering among different neutrino species. 
Since our primary goal is to have a first characterization of possible effects induced by the MAA instability, we 
will consider simple neutrino systems, whose flavor evolution is well understood in the axial symmetric case. 
We leave the study of  more complicated spectral configurations to future works.   

The structure of the paper is as follows. In Sec.~2 we present the neutrino equations of motion without 
imposing axial asymmetry. In Sec.~3 we show the results for the flavor evolution of a monochromatic beam of 
$\nu_e$ and $\overline\nu_e$ with different initial flavor asymmetries, in the presence of only MAA and both 
MAA and MZA effects.
We find that independently of the initial flavor symmetry, the MAA effects  trigger new flavor 
conversions in the (otherwise stable) NH case. These lead to a flavor decoherence of the neutrino ensemble. 
 In Sec.~3 we also include  the effect of continuous  energy spectra.
We consider  a system with an excess of  $\nu_e$ over $\overline\nu_e$, whose spectra are represented by Fermi-Dirac distributions.
Again we find new flavor conversions in NH case. Remarkably, the flavor evolution crucially depends on the asymmetry among 
 $\nu_e$ and $\overline\nu_e$. If it is sufficiently large, final spectra  show an ordered behavior with spectral swaps and splits. 
Conversely, for small flavor asymmetries  decoherence among angular modes develops, affecting the flavor evolution
also in inverted mass hierarchy (IH,  ${\Delta m^2_{\rm atm}} <0$). 
 Finally, in Sec.~4 we summarize our results and we  conclude.

\section{Equations of motion}

We work in  a two-flavor oscillation scenario, associated to the 
atmospheric mass-square difference $\Delta m^2_{\rm atm}= 2 \times 10^{-3}$~eV$^2$ and
and with the small (matter suppressed) in-medium mixing $\Theta = 10^{-3}$.  
Since we aim at isolating the effect of the MAA instability, 
we assume in the following that self-induced flavor conversions are not matter suppressed (as expected instead at $t \lesssim 1$~s
after the core bounce)~\cite{Chakraborty:2011nf,Chakraborty:2011gd,Saviano:2012yh,Sarikas:2011am}, ignoring the ordinary matter background  in the equations of motion. 

We still assume that variations in the transverse direction
are small so that the flavor evolution  depends only on $r$, $E$ and ${\bf v}_{\bf p}$.
In other words, we study neutrino propagation only in
the neighborhood of a chosen location and do not worry
about the global solution~\cite{Raffelt:2013isa}.
 We assume neutrinos to be emitted half-isotropically at the neutrinosphere, 
 that here we schematically fix at $R=10$~km. For simplicity, we neglect possible residual scatterings that could affect $\nu$'s after the 
neutrinosphere, producing a small ``neutrino halo'' that would broaden the $\nu$ angular 
distributions~\cite{Cherry:2012zw,Sarikas:2012vb}.
If we describe neutrinos by their local velocity ${\bf v}_{\bf p}$, the 
 zenith angle of a given neutrino trajectory with respect to the radial direction $\theta_r$, due to flux conservation will depend on the radial coordinate $r$ even for straight line propagation. 
It is thus convenient to parameterize every zenith angular mode in terms of its emission angle $\theta_R$ relative to the radial direction
of the neutrinosphere.  
A further simplification is obtained if one labels the different zenith angular modes in terms of the variable $u=\sin^2\theta_R$,
as in~\cite{EstebanPretel:2007ec,Banerjee:2011fj}.
  At a radius $r$, the radial velocity of a mode with angular label $u$ is
$v_{u,r} = (1-u R^2/r^2)^{1/2}$~\cite{EstebanPretel:2007ec} and the transverse velocity is 
$\beta_{u,r}= u^{1/2} R/r$~\cite{Raffelt:2013rqa}.
 Conventionally, we use negative $E$  for anti-neutrinos and we label the different energy modes
by the vacuum oscillation frequency $\omega = \Delta m^2_{\rm atm}/2E$. Then,
following~\cite{EstebanPretel:2007ec}, we define the flux matrices  ${\bf J}_{\omega,u, \varphi}$  
as function of the radial coordinate, where $\varphi$ is the azimuth angle of ${\bf v_p}$.
The diagonal ${\bf J}_{\omega,u,\varphi}$ elements are
the ordinary number fluxes $F_{\nu_{\alpha}}(\omega,u,\varphi)$ 
at a  radius $r$.  The off-diagonal elements,
which are initially zero, carry a phase information due to 
flavor mixing.
The flux matrices are represented by polarization
vectors ${\bf P}_{\omega,u, \varphi}$ in the usual way,
\begin{equation}
{\bf J}_{\omega,u, \varphi} =
\frac{\textrm{Tr}{\bf J}_{\omega,u, \varphi}}{2} +
\frac{F^R_e-F^R_x}{2}{\bf P}_{\omega,u, \varphi} \cdot \bm \sigma \,\ ,
\end{equation}
where $F^R_{e,x}(\omega, u, \varphi)$ are the flavor fluxes at the neutrinosphere  radius $R$.
Following~\cite{Raffelt:2013rqa}, we introduce the dimensionless spectrum $g(\omega, u, \varphi)$, 
representing ${F^R_e-F^R_x}$. It is negative for antineutrinos, where $\omega <0$, and normalized
to $\bar\nu$ flux, i.e. $\int_{-\infty}^{0} d \omega \int_{0}^{1} d u
\int_{0}^{2 \pi} d \varphi g(\omega, u, \varphi)=-1$. 
We introduce the $\nu-\bar\nu$ asymmetry parameter 
$\varepsilon = \int d\Gamma g$, where
$\int d \Gamma = \int_{-\infty}^{+\infty}d \omega
 \int_{0}^{1}du  \int_{0}^{2\pi}d\varphi$.
 In the following we will always assume axial symmetry of the neutrino emission and half-isotropic
 zenith angle $\nu$ distributions.
 Therefore, at the neutrinosphere 
 $ g(\omega, u, \varphi)=  g(\omega)/ 2\pi$.

We write the compact form of the 
 the equations of motion (neglecting matter effects) for the flavor-spin polarization
vectors~\cite{Banerjee:2011fj,Sigl:1992fn,Raffelt:2013rqa}\footnote{In the  flavor-spin polarization vector 
formalism, $\nu$ and $\bar\nu$ of the same flavor have polarization vectors which  point in opposite directions~\cite{Duan:2010bg}. This formalism allows
to write in a compact from the equations of motion for neutrinos and antineutrinos
[see Eq.~(\ref{eq:eom1})] . However, in our numerical examples
we prefer to use the convention in which the polarization vectors for $\nu$ and $\bar\nu$ point in the same direction.}

\begin{eqnarray}
\textrm{i}\partial_r {\bf P}_{\omega,u, \varphi} &=&
\frac{\omega {\bf B} \times {\bf P}_{\omega,u, \varphi}}{v_{u}} \nonumber\\
&+&
\mu_R \frac{R^2}{r^2} \int d \Gamma^{\prime} 
\left(\frac{1-v_{u}v_{u^\prime}-{\bm\beta}\cdot {\bm\beta}^{\prime}}
{v_{u}v_{u^\prime}} \right) 
\nonumber\\
 &\times & 
g(\omega^\prime) {\bf P}_{\omega^{\prime},u^{\prime}, \varphi^{\prime}}
\times  {\bf P}_{\omega,u, \varphi} \,\ ,
 \label{eq:eom1}
\end{eqnarray}
where  ${\bf B} =(\sin 2\Theta ,0,\pm \cos 2 \Theta)$, with
$B_z <0$ corresponds to NH, while $B_z >0$ to IH.  
The strength of the $\nu$-$\nu$ interaction is parametrized as
\begin{equation}
 \mu_R = \sqrt{2} G_F \frac{\Phi_{\bar\nu_e}(R)-\Phi_{\bar\nu_x}(R)}{4 \pi R^2} \,\ ,
 \label{eq:muR}
\end{equation}
normalized to the total ${\overline\nu}_e-{\overline\nu}_x$ flux difference at $R$. 
Finally, 
$ {\bm\beta}\cdot{\bm\beta}^{\prime} = \sqrt{u u^{\prime}}{R^2/r^2}
\cos(\varphi-\varphi^{\prime})$.
This term breaks the axial symmetry.

\begin{figure*}[!t]
 \includegraphics[angle=0,width=0.6\textwidth]{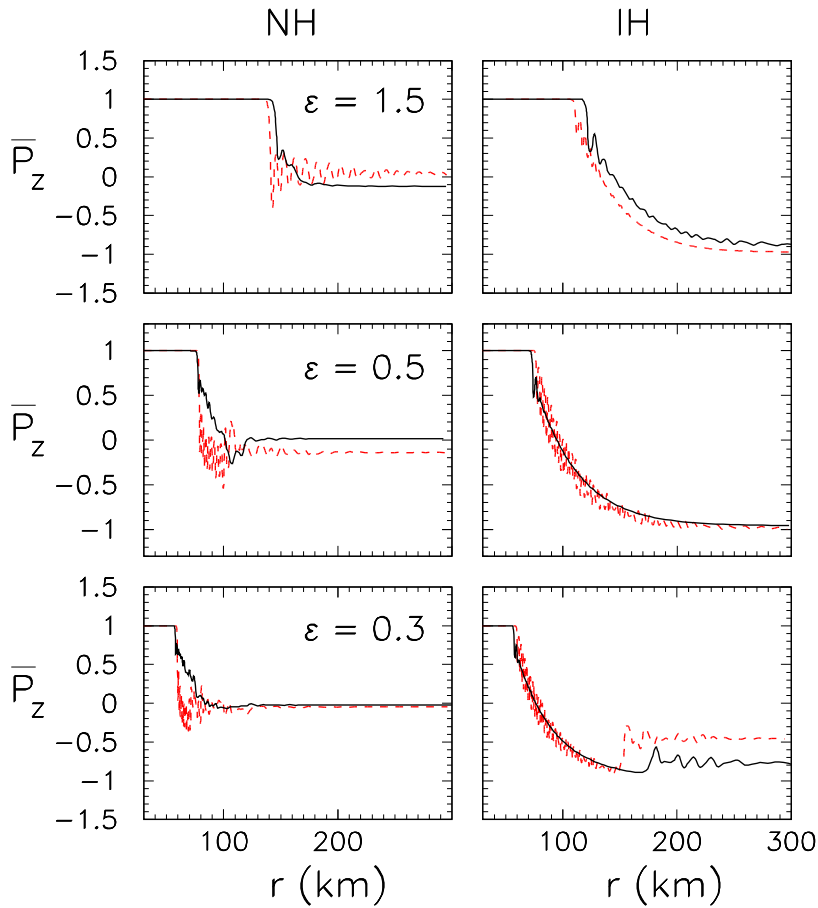} 
\caption{Single energy case with $\omega_0=0.34$ km${}^{-1}$. Radial evolution of the angle-integrated z-component 
${\bar P}_z$ of the $\bar\nu$   polarization vector  
 for different values of the flavor asymmetry $\varepsilon$.
Left panels refer to normal hierarchy, right panels to inverted hierarchy.
Dashed curves are for the MAA, SZA evolution; while
continuous curves are for the MAA, MZA cases.} \label{fig1}
\end{figure*}

We expect that self-induced flavor conversions in SNe would develop at $r \gg R$. Therefore, we can 
adopt the large-distance approximation of Eq.~(\ref{eq:eom1}) getting~\cite{Raffelt:2013rqa}
\begin{eqnarray}
\textrm{i}\partial_r {\bf P}_{\omega,u, \varphi} &=&
{\omega {\bf B} \times {\bf P}_{\omega,u, \varphi}} \nonumber \\
&-&
\mu \int d \Gamma^{\prime} 
\left[u + u^{\prime} - 2 \sqrt{u u^{\prime}} \cos(\varphi-\varphi^{\prime})  \right] \nonumber \\ 
& \times & g({\omega^\prime}) {\bf P}_{\omega^{\prime},u^{\prime}, \varphi^{\prime}}
\times  {\bf P}_{\omega,u, \varphi} \,\ ,
 \label{eq:eomld}
\end{eqnarray}
where 
\begin{equation}
\mu = \mu_R \frac{R^4}{2 r^4} \,\ .
\end{equation}

In order to numerically solve  Eq.~(\ref{eq:eomld})  we use an integration routine for stiff ordinary differential equations taken
from the NAG libraries~\cite{nag} and based on an adaptive method.

\section{Single energy neutrino beam}

In order to perform a first numerical exploration of the solution of Eq.~(\ref{eq:eomld}) 
in the presence of the  MAA effects, we consider a monochromatic $\nu$  ensemble, composed only 
by $\nu_e$ and $\bar\nu_e$  in which all the particles share
the same  energy $\langle E \rangle = 15$ MeV. This would correspond to a
vacuum oscillation frequency $\omega_0 = 0.34$ km$^{-1}$.  
We consider an excess of $\nu_e$ over $\bar\nu_e$, parametrized by the asymmetry parameter $\varepsilon$,
such that
the initial energy spectrum would be equivalent to 
$g(\omega) = -\delta (\omega+\omega_0) + (1+\varepsilon) \delta (\omega-\omega_0)$.
We fix the  $\bar\nu_e$ luminosity $L_{\bar\nu_e} = 10^{51}$ erg$/$s, so that in Eq.~(\ref{eq:muR})
\begin{eqnarray}
\mu_R &=& 0.7 \times 10^{5} \,\ \textrm{km}^{-1} \nonumber \\
&\times & 
\frac{L_{\bar\nu_e}}{\langle E\rangle}
\frac{15 \,\ \textrm{MeV}}{10^{51} \,\ \textrm{erg}/\textrm{s}}
\left(\frac{10 \,\ \textrm{km}}{R} \right)^2 \,\ .
\end{eqnarray}

The flavor evolution in this case under perfect axial symmetry is well known. 
If the flavor asymmetry $\varepsilon$ is sufficiently large (i.e. $\varepsilon \gtrsim 0.3$) the system would 
be stable in normal hierarchy, while in inverted hierarchy it would exhibit the bimodal instability leading
to complete flavor conversions in the $\bar\nu$ sector, while in the $\nu$ sector there will remain a fraction
of $\nu_e$ fixed by the conservation of the initial lepton asymmetry $\varepsilon$~\cite{Hannestad:2006nj}. 
 Conversely for smaller values of $\varepsilon$, MZA effects would lead to flavor decoherence in both
 the mass hierarchies~\cite{EstebanPretel:2007ec}.   

We show  how the flavor evolution changes including the MAA effects. 
In Fig.~\ref{fig1} we represent the radial evolution of   $z$-component
${\bar P}_z$ of the  $\bar\nu$ polarization vector integrated over $\varphi$ and $u$. In the Figure,
we normalize the integrated polarization vector such as $\bar{P}_z=+1$ corresponds to a pure  $\bar\nu_e$ system, while
$\bar{P}_z=-1$ corresponds  to all $\bar\nu_x$.
 We consider both NH (left panels) and IH (righ panels) for 
different values of the flavor asymmetry parameter, namely $\varepsilon=1.5$ (upper panels),
 $\varepsilon=0.5$ (middle panels) and  $\varepsilon=0.3$ (bottom panels). 
We compare  single-zenith-angle (SZA) evolution (dashed curves) with $u=0.5$,  with the complete MZA and MAA cases
(continuous curves).
We realize that in order to get numerical convergence of our results is enough to take ${\mathcal O}(10)$ angular modes in $\varphi$, while a  larger number of modes is required for the MZA simulations. 
We remark that numerically stable results require less modes in $\varphi$ than in $u$. 
Indeed, as discussed in~\cite{Raffelt:2013rqa} for discrete azimuthal 
angles no spurious instabilities appear, in contrast to discrete
zenith angles~\cite{Sarikas:2012ad}. 
In particular, we use $N_\varphi=30$ for $\varphi \in [0;2\pi]$ and $N_u=80$
for $u \in [0;1]$.

The most striking result of this toy model is that independently of the value of the flavor asymmetry 
$\varepsilon$, if we break the axial symmetry in the $\nu$ evolution,  NH  presents self-induced flavor conversions, as predicted from the stability 
analysis of~\cite{Raffelt:2013rqa}. 
It is interesting to mention that these flavor conversions develop even if here we are starting from an initial
perfectly spherical $\nu$ distribution. Indeed, since the system in unstable it is enough a small 
breaking of the perfect axial symmetry, induced the numerical noise: The small numerical seeds will be exponentially amplified during the flavor evolution.%
~\footnote{The speed-up of flavor instabilities under the effects of very small
seeds in the initial conditions was already pointed out in~\cite{Dasgupta:2010cd}.}
The effect of $\varepsilon$ is to shift the onset of the flavor conversions, i.e. the largest 
 $\varepsilon$ the greatest the onset radius of the flavor conversions. This relation is qualitatively
similar to what already found for the bimodal flavor evolution~\cite{Hannestad:2006nj}.  
It is also interesting to notice  that MZA effects do not seem to play an important role in changing
 the flavor dynamics. 
 The final outcome is generic: the 
MAA effect leads to a final ${\bar P}_z \simeq 0$ that would imply a flavor decoherence leading to a 
 mixture of $\bar\nu_e$ and  $\bar\nu_x$. Of course, the decoherence cannot be complete in the $\nu$
sector due to the lepton number conservation~\cite{Hannestad:2006nj}, that guarantees  a certain fraction of $\nu_e$ in the final state.  
Passing now to  IH, we find that the onset of the flavor conversions is associated with the usual
bimodal instability, leading to the well-known ``pendulum'' behaviour which tend to completely invert
${\bar P}_z$ leading to a  complete $\bar\nu_e \to \bar\nu_x$ flavor change~\cite{Hannestad:2006nj}.
For $\varepsilon=1.5, 0.5$, MAA and
MZA effects do not play a significant role, since one observes a  ``quasi-single angle''
behavior~\cite{EstebanPretel:2007ec}, driven by the bimodal instability. 
Only  for a  small flavor asymmetry $\varepsilon=0.3$, MAA effects can lead to a deviation
with respect to the expected pendulum behavior.  This can be understood as follows.
 We have found that  an ensemble initially composed of only 
$\nu_e$ and $\bar\nu_e$ is unstable  under MAA effects in  NH.
Instead,  if the initial ensemble had been consisted
by $\nu_x$ and $\bar\nu_x$,  the situation would have been
reversed, i.e. we would have found the MAA  instability for IH. 
Therefore, starting with  $\nu_e$ and $\bar\nu_e$, the MAA instability can act in IH 
\emph{only}  after bimodal flavor conversions have significantly swapped the initial fluxes
into the unstable  $\nu_x$ and $\bar\nu_x$ ones. In our numerical example, only for $\varepsilon=0.3$, 
bimodal flavor conversions start early enough to allow the MAA instability to grow and significantly affect the
 flavor evolution at large $r$.

\begin{figure*}[!t]
 \includegraphics[angle=0,width=0.4\textwidth]{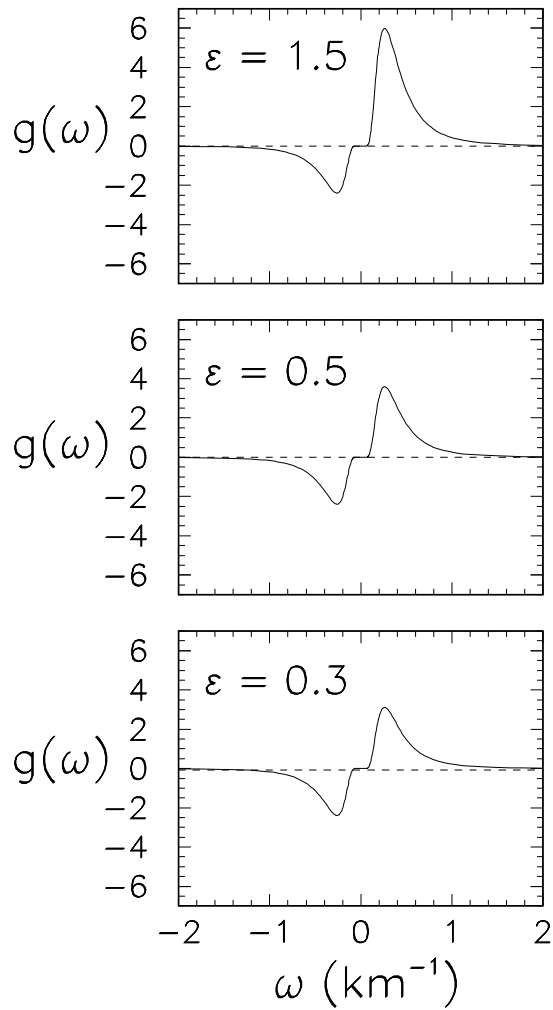} 
\caption{$g(\omega)$ spectra corresponding to $\nu_e$ and
$\bar\nu_e$ Fermi-Dirac distributions with $\langle E \rangle =15$ MeV  and  different values of the flavor asymmetry $\varepsilon$.
} \label{fig2}
\end{figure*}

\begin{figure*}[!t]
 \includegraphics[angle=0,width=0.6\textwidth]{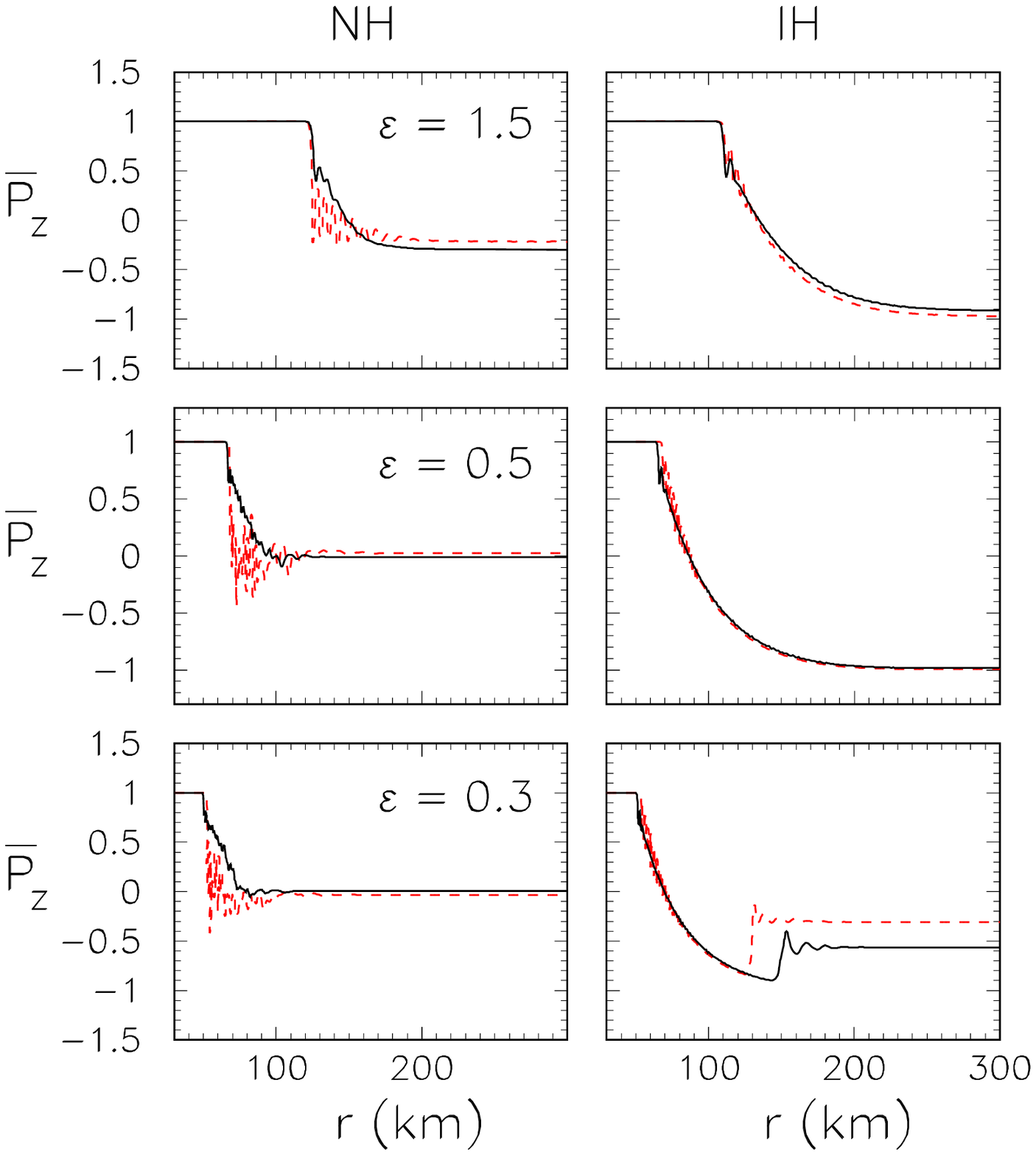} 
\caption{Multi-energy case corresponding to Fermi-Dirac energy spectra  with $\langle E \rangle =15$ MeV.
 Radial evolution of the integrated z-component 
${\bar P}_z$ of the polarization vector  for $\bar\nu$ 
 for different values of the flavor asymmetry $\varepsilon$.
Left panels refer to normal hierarchy, right panels to inverted hierarchy.
Dashed curves are for the MAA, SZA evolution; while
continuous curves are for the MAA, MZA case.} \label{fig3}
\end{figure*}

\section{Fermi-Dirac neutrino distributions}

It is known that self-induced flavor conversions can produce peculiar spectral features on 
continuous $\nu$ energy distributions, known as spectral swaps and splits~\cite{Raffelt:2007cb}. 
At this regard, we find interesting to investigate how the new  MAA instability acts 
on neutrino spectra. 
To explore this point, we take 
a system composed of only $\nu_e$ and ${\bar \nu}_e$ with an excess of $\nu_e$
parametrized in terms of the asymmetry parameter $\varepsilon$. 
This simple toy model is intended to roughly mimic
the SN $\nu$ emission, where one expects an excess of $\nu_e$ due to the core  deleptonization.  
 
We describe the $\nu$ spectra with 
Fermi-Dirac distributions with average energy $\langle E \rangle =15$ MeV and luminosity $L_{\bar\nu_e} = 10^{51}$
erg$/$s.
In Fig.~\ref{fig2} we represent the initial $g(\omega)$ spectra for our three reference values of 
 asymmetry parameter, namely $\varepsilon=1.5$ (upper panel),  $\varepsilon=0.5$ (middle panel)
 and $\varepsilon=0.3$ (lower panel). These values are in the range of the asymmetries that one would expect
 during the different phases of the SN neutrino burst. 
 
 \subsection{Flavor evolution}

Starting with the spherically symmetric $g(\omega)$ at $R$ we  evaluate the flavor evolution without axial symmetry
in both NH and IH. 
We use the same number of angular bins as in the previous Section. Moreover, we consider $N_\omega=80$ frequency modes
corresponding to a range $E\in [0;80]$~MeV. 
 In Fig.~\ref{fig3} we represent the radial evolution of the  $z$-component
${\bar P}_z$ of the $\bar\nu$ polarization vector  ${\bf P}_{\omega, u, \varphi}$
integrated over $\omega$, $\varphi$ and $u$, 
 in the same format of
Fig.~\ref{fig1}. We realize that the behavior of the  integrated ${\bar P}_z$  is very similar to what we have seen in the single-energy 
example. In particular, new flavor conversions occur in normal hierarchy triggered by the MAA instability, with 
a final  ${\bar P}_z \simeq 0$ for $\varepsilon = 0.3, 0.5$ and ${\bar P}_z \simeq -0.3$ for $\varepsilon = 1.5$.  

\begin{figure*}[!t]
 \includegraphics[angle=0,width=0.6\textwidth]{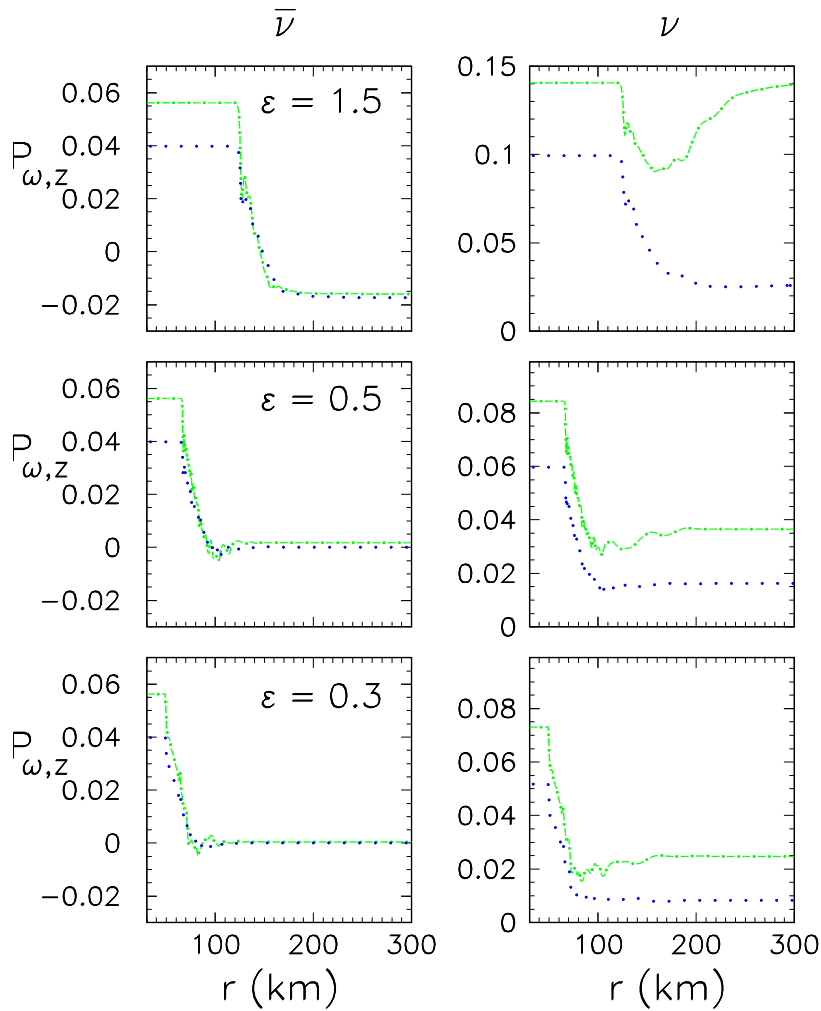} 
\caption{Normal hierarchy. Evolution of  two different $\omega$ modes  ${P}_{\omega,z}$
 for $\bar\nu_e$ (left panels) and 
$\nu_e$ (right panels)  for different values of the flavor asymmetry $\varepsilon$ in MAA, MZA case
of Fig.~\ref{fig3}.
The two lines correspond  $\omega=0.49$~km$^{-1}$ (dot-dashed curves) and
    $\omega=0.29$~km$^{-1}$ (dotted curves), respectively.
} \label{fig4}
\end{figure*}

\begin{figure*}[!t]
 \includegraphics[angle=0,width=0.6\textwidth]{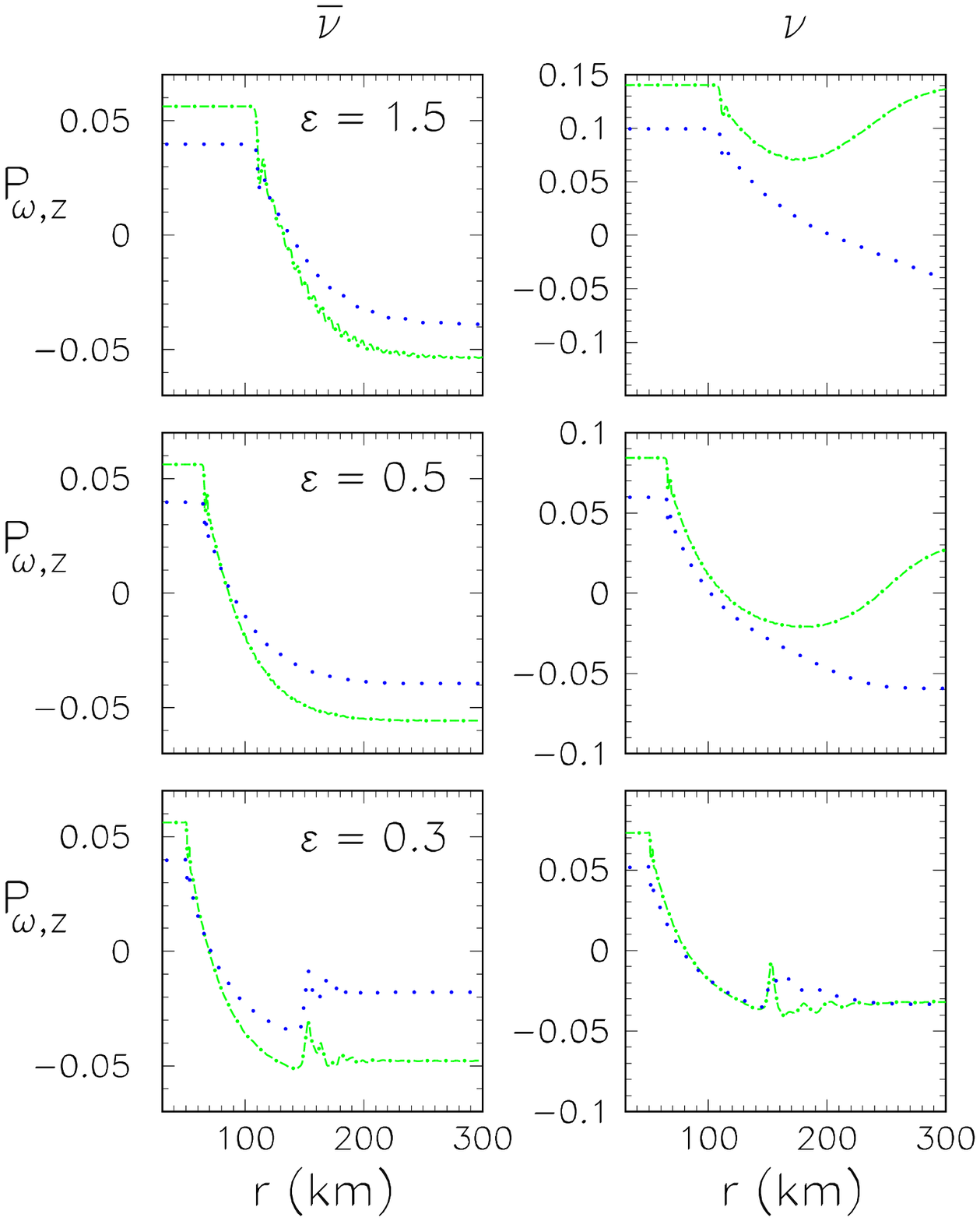} 
\caption{Inverted hierarchy. Evolution of two different $\omega$ modes  ${P}_{\omega,z}$
 for $\bar\nu_e$ (left panels) and 
$\nu_e$ (right panels)  for different values of the flavor asymmetry $\varepsilon$ in MAA, MZA case
of Fig.~\ref{fig3}.
The two lines correspond  $\omega=0.49$~km$^{-1}$ (dot-dashed curves) and
    $\omega=0.29$~km$^{-1}$ (dotted curves), respectively.
} \label{fig5}
\end{figure*}

In order to have a better understanding of how modes with different energies evolve, 
in Fig.~\ref{fig4} and~\ref{fig5} we show the radial evolution of  ${P}_{\omega,z}$ (integrated over both the angular
variables) for two representative $\omega$'s in NH
(Fig.~\ref{fig4}) and IH (Fig.~\ref{fig5}) for the same values of $\varepsilon$ of 
  Fig.~\ref{fig3} with MZA effect. In particular, left panels refer to $\bar\nu$, while right panels to $\nu$.
We have normalized the $\nu$ polarization vectors at the neutrinosphere as 
${P}_{\omega,z}= (F^R_e-F^R_x)/(F^R_e+F^R_x)$ and analogously for $\bar\nu$. 
 As energy modes,  we show  $\omega=0.49$~km$^{-1}$ (dot-dashed curves) and
    $\omega=0.29$~km$^{-1}$ (dotted curves) 
 We start discussing the NH  in  Fig.~\ref{fig4}. For large value of $\varepsilon=1.5$
 (upper panels) we realize that, 
 the ${\bar P}_{\omega,z}$  for the two different frequency modes tend to invert
 with respect to the initial value. 
 Conversely in the $\nu$ sector (right panels) the 
  $\omega=0.29$~km$^{-1}$ mode
  do invert its ${P}_{\omega,z}$ with respect to their initial value, while 
    the  $\omega=0.49$~km$^{-1}$ mode  returns to its original value at the end 
of the flavor evolution. 
  Therefore, the two different energy modes exhibit a behavior that presumably suggests the occurrence of splitting
features in the final spectra, with some parts of the spectra equal  to the original ones and other partially or completely swapped
to the other flavor.  
 Lowering the value of the asymmetry parameter $\varepsilon$ this behavior seems to change.
In particular for  $\varepsilon =0.3, 0.5$,  ${\bar P}_{\omega,z}$ tend to saturate at 0 for the two different energy modes.
In the $\nu$ sector the final ${P}_{\omega,z}$ maintain a finite final value as imposed by the conservation of the total asymmetry
$\varepsilon$. 
In our example, it seems that a larg value $\varepsilon \sim 1$
is required
to prevent flavor equilibrium induced by MAA effects.

Passing now to the IH case of Fig.~\ref{fig5}, for  $\varepsilon =1.5, 0.5$ we find the known evolution
due to the bimodal instability~\cite{Fogli:2007bk,Fogli:2008pt}. In particular, the $\bar\nu$ modes invert  ${\bar P}_{\omega,z}$ with respect 
to the initial value, while  for the $\nu$,   ${P}_{\omega,z}$ comes back to the initial value 
  for  $\omega=0.49$~km$^{-1}$ mode
   and invert their value for the other lower frequency mode. 
This behavior would produce a spectral split in the $\nu$ sector and an almost complete swap in the  $\bar\nu$ channel.
 Finally, if we consider the case with   $\varepsilon=0.3$ we realize that   MAA instability  produces 
 a modification in the flavor evolution around $r \simeq 140$~km, disturbing the expected dynamics associated
with the bimodal evolution.
 As a consequence one can foresee  a smearing of the splitting features produced for larger
$\varepsilon$.

\subsection{Oscillated spectra}

\begin{figure*}[!t]
 \includegraphics[angle=0,width=0.6\textwidth]{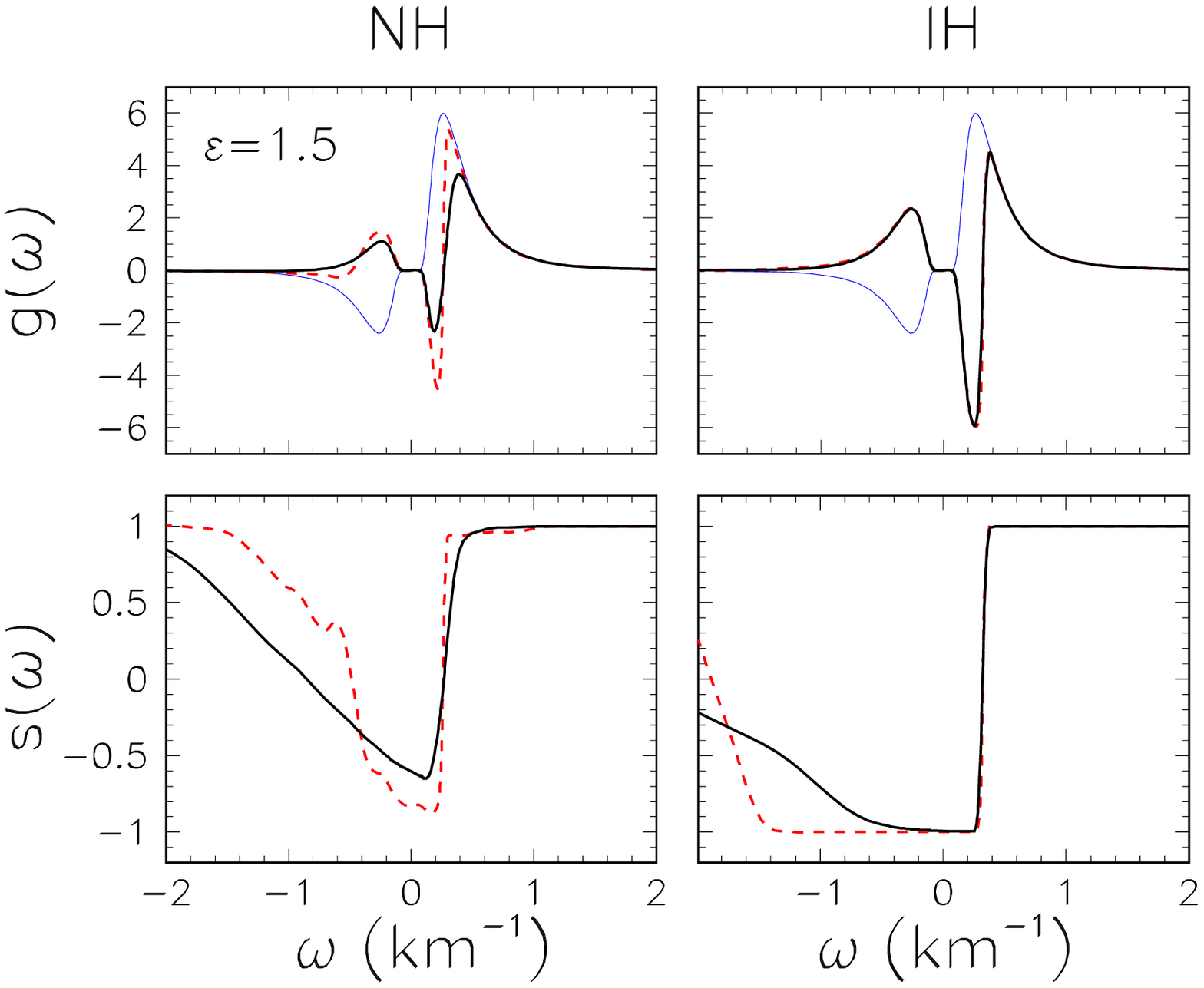} 
\caption{Asymmetry parameter $\varepsilon=1.5$. Upper panels: initial  (thin continuous curves) and final $\nu$ spectrum
for  the MAA evolution
in SZA (dashed curves) and MZA (thick continuous curves) case
 for normal hierarchy (left panel) and inverted hierarchy (right panel).
Lower panels: swap function, i.e. ratio of final with initial spectra 
in SZA  (dashed curves) and MZA (thick continuous curves).} \label{fig9}
\end{figure*}

\begin{figure*}[!t]
 \includegraphics[angle=0,width=0.6\textwidth]{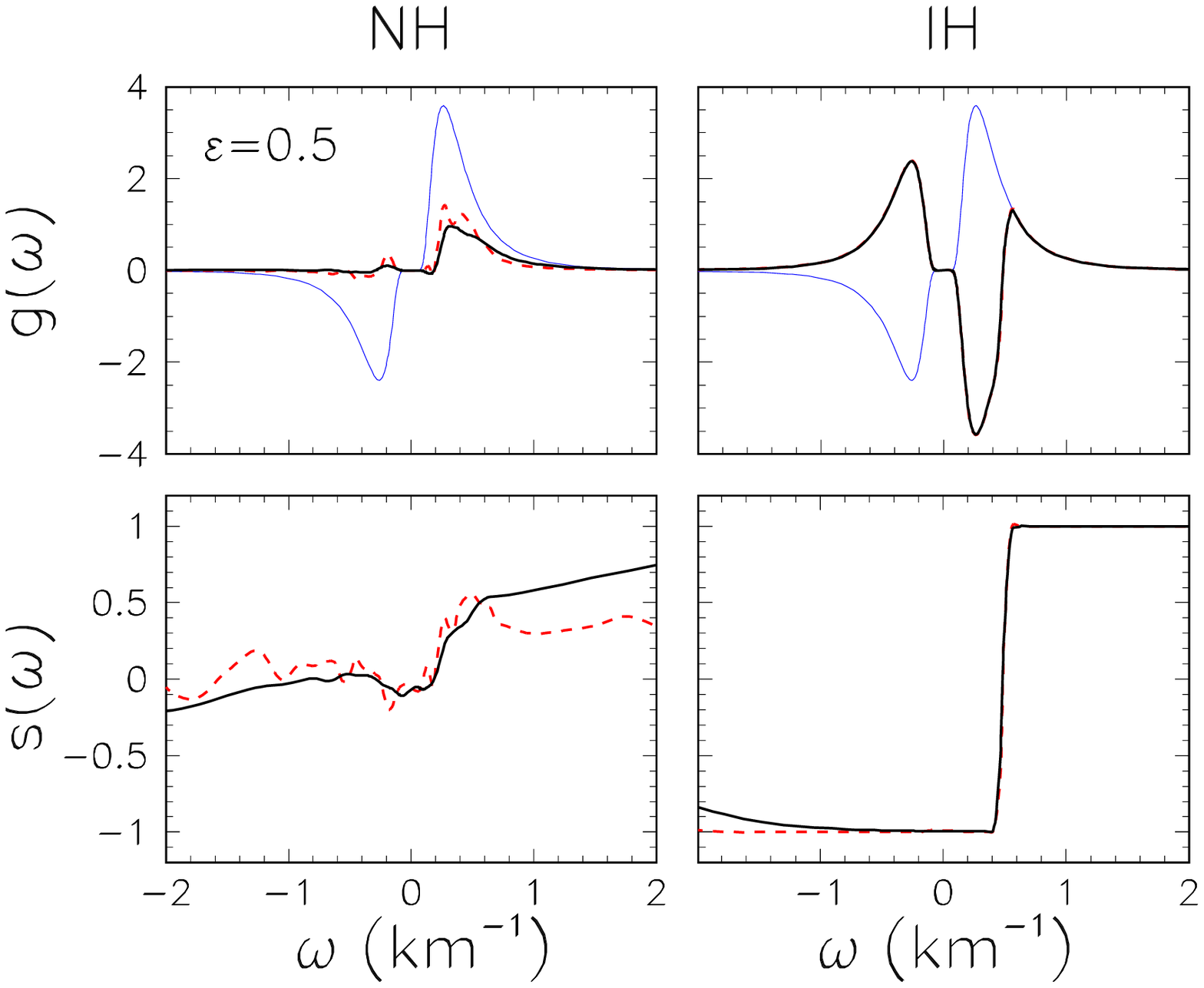} 
\caption{Asymmetry parameter $\varepsilon=0.5$. Upper panels: initial  (thin continuous curves) and final $\nu$ spectrum
for  the MAA evolution
in SZA (dashed curves) and MZA (thick continuous curves) case
 for normal hierarchy (left panel) and inverted hierarchy (right panel).
Lower panels: swap function, i.e. ratio of final with initial spectra  
in SZA  (dashed curves) and MZA (thick continuous curves).} \label{fig10}
\end{figure*}

\begin{figure*}[!t]
 \includegraphics[angle=0,width=0.6\textwidth]{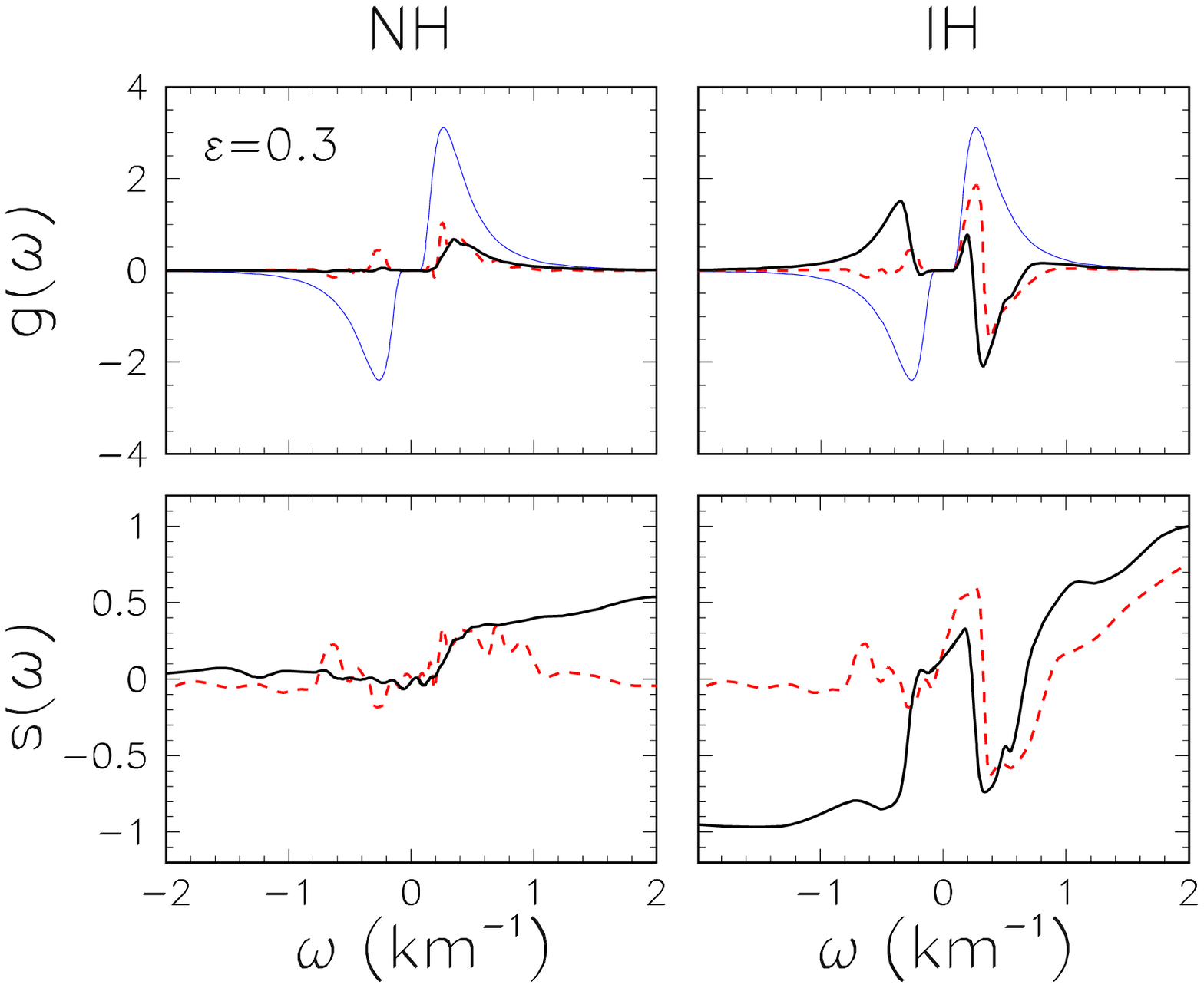} 
\caption{Asymmetry parameter $\varepsilon=0.3$. Upper panels: initial  (thin continuous curves) and final $\nu$ spectrum
for  the MAA evolution
in SZA (dashed curves) and MZA (thick continuous curves) case
 for normal hierarchy (left panel) and inverted hierarchy (right panel).
Lower panels: swap function, i.e. ratio of final with initial spectra 
in SZA  (dashed curves) and MZA (thick continuous curves).} \label{fig11}
\end{figure*}

In order to show how this flavor dynamics triggered by the MAA effects impacts the oscillated $\nu$ spectra,
in the following three Figures we represent  in the upper panels the initial $g(\omega)$ (light continuous curves),
the final one under MAA flavor evolution with SZA effect (dashed curve) and the final one with MZA effect (thick continuous curves).
In the lower panels we represent the swap function $s(\omega) = g^{\textrm{final}}(\omega)/ g^{\textrm{initial}}(\omega)$
for the SZA evolution (dashed curves) and  the MZA case (thick continuous curves).
Left panels refer to  NH , while right panels to IH. 
In particular, Fig.~\ref{fig9} is for $\varepsilon=1.5$, 
  Fig.~\ref{fig10} for $\varepsilon=0.5$ and Fig.~\ref{fig11} for $\varepsilon=0.3$. 

Starting with $\varepsilon=1.5$ in  Fig.~\ref{fig9}, we realize that the final spectra in NH and IH are qualitatively similar. 
 The MAA instability in NH produces a final 
 spectrum with a split at $\omega \simeq 0.2$~km$^{-1}$, similar to the one   induced in IH by
the bimodal instability at $\omega \simeq 0.4$~km$^{-1}$.  
 In the SZA evolution another split would appear in the $\bar\nu$ at $\omega \simeq -0.5$~km$^{-1}$ in NH,
 and at $\omega \simeq -1.5$~km$^{-1}$ in IH. However, both these $\bar\nu$ splits are rather fragile under the
 MZA effects, which tend to smear them. 

In Fig.~\ref{fig10} we consider  $\varepsilon=0.5$. We realize that  IH evolution  is very similar to the 
one observed for $\varepsilon=1.5$, apart from the disappearance  of the $\bar\nu$ split. Conversely, in NH
the final spectrum is remarkably different. From the swap function $s(\omega)$ one observes that the 
swaps and splits are significantly smeared in both  SZA and MZA situations. The swap function for $\bar\nu$
tends to 
flatten, approaching  zero in the $\bar\nu$ sector, indicating the tendency towards flavor decoherence.

Finally in Fig.~\ref{fig11} we consider  $\varepsilon=0.3$. In NH the flavor decoherence is complete
with $s(\omega) \sim 0$ for $\bar\nu$. For $\nu$ for $\omega > 0.5$~km$^{-1}$ one finds 
 $s(\omega) \sim 0$  in SZA ,  while  $s(\omega) \sim 0.5$ in  MZA.
  For smaller $\omega$ the swap function tends towards zero in both 
 the situations.
In IH, the MAA effects produce a strong smearing of the splitting features.
 Multiple splits appear along the spectrum. In the $\bar\nu$ part the SZA and MZA cases exhibit a different
 behavior for  $\omega < -0.4$~km$^{-1}$.  In particular, $s(\omega) \simeq 0$ in the SZA scheme, while
 $s(\omega) \simeq -1$ in the MZA evolution.

From these results, one can conclude that the new found MAA instability can lead to a rich phenomenology. The general 
outcome of the flavor evolution is not a trivial flavor decoherence as one could have expected from the single-energy evolution. 
Conversely, an ordered behavior can be present in the final spectra. As for the MZA effect, a crucial parameter
to suppress the decoherence is the flavor asymmetry parameter $\varepsilon$. 
However, at least from our numerical examples,
it seems that a large value of   $\varepsilon$ (i.e. $\varepsilon\sim 1$) is required to suppress the MAA decoherence, while 
it was found that 
a smaller value, i.e. 
$\varepsilon \gtrsim 0.3$,  is enough to suppress the MZA instability~\cite{EstebanPretel:2007ec}.

\subsection{Multipoles evolution}

\begin{figure*}[!t]
 \includegraphics[angle=0,width=0.6\textwidth]{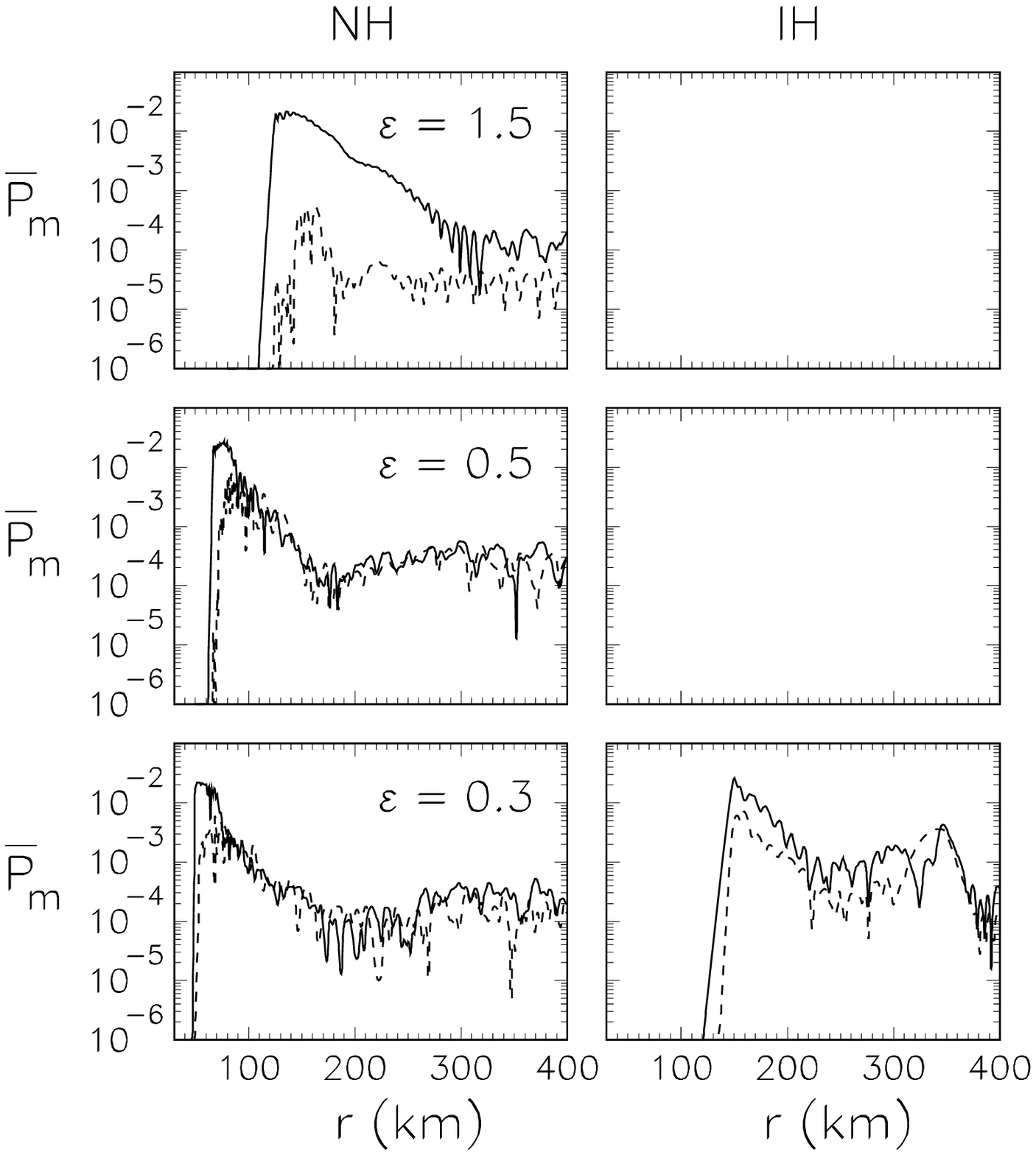} 
\caption{Radial evolution of the multipoles $P_m$ for $\bar\nu$ in the MAA evolution for the  MZA case
of Fig.~\ref{fig3}.
Left panels refer to normal hierarchy, right panels to inverted hierarchy.
Continuous curves refer to $m=1$, dashed ones to $m=2$. 
} \label{fig6}
\end{figure*}
 
 When the axial symmetry is broken  in the flavor evolution, the polarization vectors can develop significant multipoles
in $\cos\varphi$ or $\sin\varphi$. At this regard,
in order to compare the development of MAA effects in the different cases we studied, we find interesting
to show the development of the multipole expansion of the polarization vectors in $\cos\varphi$.
  
Since the onset of the MAA instability is related to an exponential growth of the transverse components $(x,y)$ of the 
polarization vector, we define
\begin{equation}
P^{x,y}_m = \frac{1}{2\pi} \int d\Gamma \,\   g(\omega) P^{x,y}_{\omega,u, \varphi} \cos m\varphi \,\ ,
\end{equation}
and
\begin{equation}
P_m = \sqrt{(P^x_m)^2 + (P^y_m)^2} \,\ , 
\end{equation}
where $m=1,2,3 \ldots$. 
In Fig.~\ref{fig6} we represent the first two multipoles $P_m$ for $\bar\nu$ for  the MAA evolution in the MZA case
shown in Fig.~\ref{fig3} using the same format of that Figure.  
In particular, the continuous curves refer to the dipole $m=1$, while the dashed one to $m=2$. 
 Starting from $\varepsilon=1.5$ we see that in NH $P_1$ rises exponentially to $\sim 10^{-2}$ 
 at $r\simeq 100$~km corresponding to the onset of the flavor conversions. Then, it declines by roughly
 one order of magnitude at $r\simeq 200$~km. It is interesting to realize that the $m=2$  multipole
at its peak is suppressed by roughly two orders of magnitude with respect to $P_1$.
In  IH, no multipole develops since the polarization
vectors maintain a spherically symmetric distribution. 
 Passing now to the case of $\varepsilon=0.5$  we realize that the $P_m$ rises at $r\simeq 70$~km.
In comparison with the previous case we see that   $P_{2}$ 
at the peak is only a factor $\sim 3$ smaller than  $P_1$. Moreover, 
 at $r\gtrsim 100$~km    the two different multipoles have a comparable strength. 
 This seems an important difference with respect to the $\varepsilon=1.5$ case and corresponds to
a different flavor evolution. In particular, for  $\varepsilon=1.5$ where a rather ordered behavior 
 in the polarization vector occurs, only the dipole is significantly excited.
Conversely, for  $\varepsilon=0.5$  where the $\nu$ system tends to evolve towards the flavor decoherence, 
also higher order multipoles are significantly excited. 
 In this case in IH again one does not observe any development of multipoles. 
 
 Finally, for $\varepsilon=0.3$  the NH has a behavior  similar to the previous one, with the different multipoles 
 significantly excited at $r \simeq 50$~km. 
 Moreover, in IH the different multipoles get excited at $r\simeq 140$~km, where the MAA instability starts to develop 
after the bimodal instability swapped the original spectra.

\section{Conclusions}

Very recently it has been pointed out that removing the axial symmetry in the solution
of the non-linear evolution equations of SN neutrinos, a new  multi-azimuthal-angle instability would
emerge~\cite{Raffelt:2013rqa,Raffelt:2013isa}. Inspired by this important finding, we performed the first simulations
of the SN neutrino flavor evolution including the azimuthal angle $\varphi$ as explicit variable. 
In order to understand this flavor dynamics, 
we considered simple toy models in which we prepared our system initially with   
 an excess of  $\nu_e$ over $\bar\nu_e$. 
 We started considering monochromatic neutrino beams. 
 In this case, in NH significant flavor conversions are triggered by the MAA instability generically leading  to
 flavor decoherence of the neutrino ensemble. 
 In IH the MAA instability can affect the flavor evolution only for small neutrino asymmetries, but only after the usual bimodal
 instability has inverted the initial flavor content. 
 Considering more realistic continuous energy distributions, described by Fermi-Dirac energy spectra, we find that the MAA instability can lead
to a rich phenomenology. Indeed, the flavor decoherence found in the single-energy case is not the unique possibility. 
In particular, for large flavor asymmetries, oscillated spectra exhibit an ordered behavior with swaps and splits, similar to the ones
produces in IH by the bimodal instability. 
Only for small flavor asymmetries the flavor decoherence is unavoidable. 

This result observed with our simple examples
would presumably  lead   to a change of paradigm in the characterization of self-induced SN neutrino oscillations. 
In order to assess the impact of the MMA instability on the flavor evolution of SN neutrinos, 
one necessarily would move from  toy models to a more 
realistic characterization of the dense SN neutrino gas. In particular, it is well-known that
the self-induced effects are crucially dependent on the
flux ordering among different neutrino species (see, e.g.,~\cite{Mirizzi:2010uz}). 
Moreover, the role of dense ordinary matter, that could suppress the self-induced effects during the 
accretion phase~\cite{Chakraborty:2011nf,Chakraborty:2011gd,Saviano:2012yh,Sarikas:2011am}
needs to be explored. Indeed, according to the stability analysis performed in~\cite{Raffelt:2013rqa}, in the presence of MAA effects
 matter suppression would
require larger densities than previously thought.
Furthermore, in the present study we have assumed perfect spherical neutrino emission, which does not occur in multi-dimensional
SN simulations~\cite{Ott:2008jb,Tamborra:2013laa}. In future works it would be interesting to understand how the removal of the initial spherical symmetry would
affect the further flavor evolution. 

In conclusion, the discovery of these new effects adds many open questions and additional
layers of complications in the simulation of the flavor
evolution for supernova neutrinos. 
At the moment previous  results of the self-induced effects on SN neutrinos, based on the unjustified assumption
of   axial symmetry 
could not be taken as granted, but should be reexamined. 
For sure, further studies are necessary to characterize 
this fascinating flavor dynamics and to get a deeper theoretical understanding of its effects. At this regard, the possibility to
detect signatures of self-induced oscillations in the next galactic
supernova neutrino burst~\cite{Choubey:2010up}  would motivate further
analytical and numerical investigations.

\section*{Acknowledgements} 

A.M. kindly thanks Georg Raffelt for many interesting discussions and suggestions
during the development of this work.
Moreover, he acknowledges Basudeb Dasgupta, Huaiyu Duan and Irene Tamborra  for interesting comments on the manuscript.
This work  was supported by the German Science Foundation (DFG)
within the Collaborative Research Center 676 ``Particles, Strings and the
Early Universe.''



\begin{thebibliography}{00}

\bibitem{Raffelt:2013rqa} 
  G.~Raffelt, S.~Sarikas and D.~d.~S.~Seixas,
  Phys.\ Rev.\ Lett.\  {\bf 111}, 091101 (2013)
  [arXiv:1305.7140 [hep-ph]].


  
 \bibitem{Matt}  L.~Wolfenstein,  
				``Neutrino Oscillations In Matter,''  
                Phys.\ Rev.\ D {\bf 17}, 2369 (1978);  
                S. P.~Mikheev and A. Yu.\ Smirnov,  
                ``Resonance Enhancement Of Oscillations In Matter And Solar Neutrino  
				Spectroscopy,''  
                Yad.\ Fiz.\ {\bf 42}, 1441 (1985)  
                [Sov.\ J.\ Nucl.\ Phys.\ {\bf 42}, 913 (1985)].  



\bibitem{Dighe:1999bi}
  A.~S.~Dighe and A.~Y.~Smirnov,
  ``Identifying the neutrino mass spectrum from the neutrino burst from a
  supernova,''
  Phys.\ Rev.\  D {\bf 62}, 033007 (2000)
  [hep-ph/9907423].
  
  
\bibitem{Pantaleone:1992eq}  
  J.~Pantaleone,  
  ``Neutrino oscillations at high densities,''  
  Phys.\ Lett.\ B {\bf 287}, 128 (1992).  
  
\bibitem{Sigl:1992fn}  
  G.~Sigl and G.~Raffelt,  
  ``General kinetic description of relativistic mixed neutrinos,''  
  Nucl.\ Phys.\ B {\bf 406}, 423 (1993).  
  
    
\bibitem{McKellar:1992ja}
  B.~H.~J.~McKellar and M.~J.~Thomson,
  ``Oscillating doublet neutrinos in the early universe,''
  Phys.\ Rev.\  D {\bf 49}, 2710 (1994).
  
  
\bibitem{Qian:1994wh}
  Y.~Z.~Qian and G.~M.~Fuller,
  ``Neutrino-neutrino scattering and matter enhanced neutrino flavor
  transformation in Supernovae,''
  Phys.\ Rev.\  D {\bf 51}, 1479 (1995)
  [astro-ph/9406073].
  
\bibitem{Samuel:1993uw}  
  S.~Samuel,  
  ``Neutrino oscillations in dense neutrino gas\-es,''  
  Phys.\ Rev.\ D {\bf 48}, 1462 (1993).  
  
\bibitem{Kostelecky:1993dm}  
  V.~A.~Kosteleck\'y and S.~Samuel,  
  ``Neutrino oscillations in the early universe with an inverted  
  neutrino mass hierarchy,''  
  Phys.\ Lett.\ B {\bf 318}, 127 (1993).  
  
\bibitem{Kostelecky:1995dt}  
  V.~A.~Kosteleck\'y and S.~Samuel,  
  ``Self-maintained coherent oscillations in dense neutrino gases,''  
  Phys.\ Rev.\ D {\bf 52}, 621 (1995)  
  [hep-ph/9506262].  
  
\bibitem{Samuel:1996ri}  
  S.~Samuel,  
  ``Bimodal coherence in dense selfinteracting neutrino gases,''  
  Phys.\ Rev.\ D {\bf 53}, 5382 (1996)  
  [hep-ph/9604341].  



  
\bibitem{Pastor:2001iu}  
  S.~Pastor, G.~G.~Raffelt and D.~V.~Semikoz,  
  ``Physics of synchronized neutrino oscillations caused by  
  self-interactions,''  
  Phys.\ Rev.\ D {\bf 65}, 053011 (2002)  
  [hep-ph/0109035].  
 
\bibitem{Pastor:2002we}  
  S.~Pastor and G.~Raffelt,  
  ``Flavor oscillations in the supernova hot bubble region:  
  Nonlinear  effects of neutrino background,''  
  Phys.\ Rev.\ Lett.\  {\bf 89}, 191101 (2002)  
  [astro-ph/0207281].  

  
\bibitem{Sawyer:2005jk}  
  R.~F.~Sawyer,  
  ``Speed-up of neutrino transformations in a supernova environment,''  
  Phys.\ Rev.\  D {\bf 72}, 045003 (2005)  
  [hep-ph/0503013].  


\bibitem{Duan:2005cp}
  H.~Duan, G.~M.~Fuller and Y.~Z.~Qian,
  ``Collective Neutrino Flavor Transformation In Supernovae,''
  Phys.\ Rev.\  D {\bf 74}, 123004 (2006)
  [astro-ph/0511275].


\bibitem{Duan:2006an}
  H.~Duan, G.~M.~Fuller, J.~Carlson and Y.~Z.~Qian,
  ``Simulation of coherent non-linear neutrino flavor transformation in the
  supernova environment. I: Correlated neutrino trajectories,''
  Phys.\ Rev.\  D {\bf 74}, 105014 (2006)
  [astro-ph/0606616].
  
\bibitem{Hannestad:2006nj}
  S.~Hannestad, G.~G.~Raffelt, G.~Sigl and Y.~Y.~Y.~Wong,
  ``Self-induced conversion in dense neutrino gases: Pendulum in flavour  
 space,''
  Phys.\ Rev.\  D {\bf 74}, 105010  (2006)
  [Erratum-ibid.\  D {\bf 76},  029901 (2007)]
  [astro-ph/0608695].


\bibitem{Duan:2010bg} 
  H.~Duan, G.~M.~Fuller and Y.~-Z.~Qian,
  ``Collective Neutrino Oscillations,''
  Ann.\ Rev.\ Nucl.\ Part.\ Sci.\  {\bf 60}, 569 (2010)
  [arXiv:1001.2799 [hep-ph]].



\bibitem{Fogli:2007bk}
  G.~L.~Fogli, E.~Lisi, A.~Marrone and A.~Mirizzi,
  ``Collective neutrino flavor transitions in supernovae and the role of
  trajectory averaging,''
  JCAP {\bf 0712}, 010 (2007)
  [arXiv:0707.1998 [hep-ph]].

\bibitem{Fogli:2008pt}
  G.~L.~Fogli, E.~Lisi, A.~Marrone, A.~Mirizzi and I.~Tamborra,
  ``Low-energy spectral features of supernova (anti)neutrinos in inverted
  hierarchy,''
  Phys.\ Rev.\  D {\bf 78}, 097301 (2008)
  [arXiv:0808.0807 [hep-ph]].



\bibitem{Raffelt:2007cb}
  G.~G.~Raffelt and A.~Y.~Smirnov,
  ``Self-induced spectral splits in supernova neutrino fluxes,''
  Phys.\ Rev.\  D {\bf 76}, 081301 (2007)
  [Erratum-ibid.\  D {\bf 77}, 029903 (2008)]
  [arXiv:0705.1830 [hep-ph]].


\bibitem{Raffelt:2007xt}
  G.~G.~Raffelt, A.~Y.~Smirnov,
  ``Adiabaticity and spectral splits in collective neutrino transformations,''
  Phys.\ Rev.\  {\bf D76}, 125008 (2007).
  [arXiv:0709.4641 [hep-ph]].

\bibitem{Duan:2007bt}
  H.~Duan, G.~M.~Fuller, J.~Carlson {\it et al.},
  ``Neutrino Mass Hierarchy and Stepwise Spectral Swapping of Supernova Neutrino Flavors,''
  Phys.\ Rev.\ Lett.\  {\bf 99}, 241802 (2007).
  [arXiv:0707.0290 [astro-ph]].


\bibitem{Duan:2008za}
  H.~Duan, G.~M.~Fuller, Y.~-Z.~Qian,
  ``Stepwise spectral swapping with three neutrino flavors,''
  Phys.\ Rev.\  {\bf D77}, 085016 (2008).
  [arXiv:0801.1363 [hep-ph]].


\bibitem{Gava:2008rp}
  J.~Gava and C.~Volpe,
  ``Collective neutrinos oscillation in matter and CP-violation,''
  Phys.\ Rev.\  D {\bf 78}, 083007 (2008)
  [arXiv:0807.3418 [astro-ph]].


\bibitem{Gava:2009pj}
  J.~Gava, J.~Kneller, C.~Volpe and G.~C.~McLaughlin,
  ``A dynamical collective calculation of supernova neutrino signals,''
  Phys.\ Rev.\ Lett.\  {\bf 103}, 071101 (2009)
  [arXiv:0902.0317 [hep-ph]].


\bibitem{Dasgupta:2009mg}
  B.~Dasgupta, A.~Dighe, G.~G.~Raffelt and A.~Y.~Smirnov,
  ``Multiple Spectral Splits of Supernova Neutrinos,''
  Phys.\ Rev.\ Lett.\  {\bf 103}, 051105 (2009)
  [arXiv:0904.3542 [hep-ph]].




\bibitem{Friedland:2010sc}
  A.~Friedland,
  ``Self-refraction of supernova neutrinos: mixed spectra and three-flavor
  instabilities,''
  Phys.\ Rev.\ Lett.\  {\bf 104}, 191102 (2010)
  [arXiv:1001.0996 [hep-ph]].

\bibitem{Dasgupta:2010cd}
  B.~Dasgupta, A.~Mirizzi, I.~Tamborra and R.~ Tom{\`a}s,
  ``Neutrino mass hierarchy and three-flavor spectral splits of supernova
  neutrinos,''
  Phys.\ Rev.\  D {\bf 81}, 093008 (2010)
  [arXiv:1002.2943 [hep-ph]].

\bibitem{Mirizzi:2010uz} 
  A.~Mirizzi and R.~ Tom{\`a}s,
  ``Multi-angle effects in self-induced oscillations for different supernova neutrino fluxes,''
  Phys.\ Rev.\ D {\bf 84}, 033013 (2011)
  [arXiv:1012.1339 [hep-ph]].


\bibitem{Pantaleone:1992xh}
  J.~T.~Pantaleone,
  ``Dirac neutrinos in dense matter,''
  Phys.\ Rev.\  D {\bf 46}, 510 (1992).



\bibitem{Raffelt:2007yz}
  G.~G.~Raffelt and G.~Sigl,
  ``Self-induced decoherence in dense neutrino gases,''
  Phys.\ Rev.\  D {\bf 75}, 083002 (2007)
  [hep-ph/0701182].
  
  
  
\bibitem{EstebanPretel:2007ec}
  A.~Esteban-Pretel, S.~Pastor, R.~ Tom{\`a}s, G.~G.~Raffelt and G.~Sigl,
  ``Decoherence in supernova neutrino transformations suppressed by
  deleptonization,''
  Phys.\ Rev.\  D {\bf 76}, 125018 (2007)
  [arXiv:0706.2498 [astro-ph]].

\bibitem{Sawyer:2008zs}
  R.~F.~Sawyer,
  ``The multi-angle instability in dense neutrino systems,''
  Phys.\ Rev.\  D {\bf 79} (2009) 105003
  [arXiv:0803.4319 [astro-ph]].


\bibitem{Mirizzi:2011tu} 
  A.~Mirizzi and P.~D.~Serpico,
  ``Instability in the Dense Supernova Neutrino Gas with Flavor-Dependent Angular Distributions,''
  Phys.\ Rev.\ Lett.\  {\bf 108}, 231102 (2012)
  [arXiv:1110.0022 [hep-ph]].


\bibitem{Mirizzi:2012wp} 
  A.~Mirizzi and P.~D.~Serpico,
  ``Flavor Stability Analysis of Dense Supernova Neutrinos with Flavor-Dependent Angular Distributions,''
  Phys.\ Rev.\ D {\bf 86}, 085010 (2012)
  [arXiv:1208.0157 [hep-ph]].

\bibitem{Banerjee:2011fj} 
  A.~Banerjee, A.~Dighe and G.~Raffelt,
  ``Linearized flavor-stability analysis of dense neutrino streams,''
  Phys.\ Rev.\ D {\bf 84}, 053013 (2011)
  [arXiv:1107.2308 [hep-ph]].

\cite{Raffelt:2013isa}
\bibitem{Raffelt:2013isa} 
  G.~Raffelt and D.~d.~S.~Seixas,
  Phys.\  Rev.\  D 88, {\bf 045031} (2013)
  [arXiv:1307.7625 [hep-ph]].


\bibitem{Chakraborty:2011nf} 
  S.~Chakraborty, T.~Fischer, A.~Mirizzi, N.~Saviano and R.~Tomas,
  ``No collective neutrino flavor conversions during the supernova accretion phase,''
  Phys.\ Rev.\ Lett.\  {\bf 107}, 151101 (2011)
  [arXiv:1104.4031 [hep-ph]].
  
\bibitem{Chakraborty:2011gd} 
  S.~Chakraborty, T.~Fischer, A.~Mirizzi, N.~Saviano and R.~Tomas,
  ``Analysis of matter suppression in collective neutrino oscillations during the supernova accretion phase,''
  Phys.\ Rev.\ D {\bf 84}, 025002 (2011)
  [arXiv:1105.1130 [hep-ph]].
  
\bibitem{Saviano:2012yh} 
  N.~Saviano, S.~Chakraborty, T.~Fischer and A.~Mirizzi,
  ``Stability analysis of collective neutrino oscillations in the supernova accretion phase with realistic energy and angle distributions,''
  Phys.\ Rev.\ D {\bf 85}, 113002 (2012)
  [arXiv:1203.1484 [hep-ph]].
  
\bibitem{Sarikas:2011am} 
  S.~Sarikas, G.~G.~Raffelt, L.~Hudepohl and H.~-T.~Janka,
  ``Suppression of Self-Induced Flavor Conversion in the Supernova Accretion Phase,''
  Phys.\ Rev.\ Lett.\  {\bf 108}, 061101 (2012)
  [arXiv:1109.3601 [astro-ph.SR]].
  
\bibitem{Cherry:2012zw} 
  J.~F.~Cherry, J.~Carlson, A.~Friedland, G.~M.~Fuller and A.~Vlasenko,
  ``Neutrino scattering and flavor transformation in supernovae,''
  Phys.\ Rev.\ Lett.\  {\bf 108}, 261104 (2012)
  [arXiv:1203.1607 [hep-ph]].
  
\bibitem{Sarikas:2012vb} 
  S.~Sarikas, I.~Tamborra, G.~Raffelt, L.~Hudepohl and H.~-T.~Janka,
  ``Supernova neutrino halo and the suppression of self-induced flavor conversion,''
  Phys.\ Rev.\ D {\bf 85}, 113007 (2012)
  [arXiv:1204.0971 [hep-ph]].
  
  

 \bibitem{nag} 
{\tt http://www.nag.com/numeric/fl/manual/html
/FLlibrarymanual.asp}

\bibitem{Sarikas:2012ad} 
  S.~Sarikas, D.~d.~S.~Seixas and G.~Raffelt,
  ``Spurious instabilities in multi-angle simulations of collective flavor conversion,''
  Phys.\ Rev.\ D {\bf 86}, 125020 (2012)
  [arXiv:1210.4557 [hep-ph]].
    
\bibitem{Dasgupta:2010cd}
  B.~Dasgupta, A.~Mirizzi, I.~Tamborra and R.~Tomas,
 ``Neutrino mass hierarchy and three-flavor spectral splits of supernova neutrinos,''
  Phys.\ Rev.\ D {\bf 81} (2010) 093008
  [arXiv:1002.2943 [hep-ph]].
    
\bibitem{Choubey:2010up} 
  S.~Choubey, B.~Dasgupta, A.~Dighe and A.~Mirizzi,
  ``Signatures of collective and matter effects on supernova neutrinos at large detectors,''
  arXiv:1008.0308 [hep-ph].
  
\bibitem{Ott:2008jb} 
  C.~D.~Ott, A.~Burrows, L.~Dessart and E.~Livne,
  ``2D Multi-Angle, Multi-Group Neutrino Radiation-Hydrodynamic Simulations of Postbounce Supernova Cores,''
  Astrophys.\ J.\  {\bf 685}, 1069 (2008)
  [arXiv:0804.0239 [astro-ph]].
 
\bibitem{Tamborra:2013laa} 
  I.~Tamborra, F.~Hanke, B.~Mueller, H.~-T.~Janka and G.~Raffelt,
  ``Neutrino signature of supernova hydrodynamical instabilities in three dimensions,''
  Phys.\  Rev.\  Lett.\  {\bf 111}, 121104 (2013)
  [arXiv:1307.7936 [astro-ph.SR]].
 
  
 \end{thebibliography}
\end{document}